\documentclass[]{elsart}

\usepackage{graphicx}
\usepackage{amsmath}
\usepackage{amssymb}
\usepackage{bm}

\begin{document}

\begin{frontmatter}

\journal{Physics Letters B}

\title{Left-right asymmetry in semi-inclusive deep inelastic
scattering process}

\author{Jun She},
\author{Yajun Mao},
\author{Bo-Qiang Ma}\ead{mabq@phy.pku.edu.cn}
\address{School of Physics and State Key Laboratory of Nuclear
Physics and Technology, Peking University, Beijing 100871, China}

\begin{abstract}
We analyze the left-right asymmetry of pion production in
semi-inclusive deep inelastic scattering~(SIDIS) process of
unpolarized charged lepton on transversely polarized nucleon target.
Unlike available treatments, in which some specific weighting
functions are multiplied to separate theoretically motivated
quantities, we do not introduce any weighting function following the
analyzing method by the E704 experiment. The advantage is that this
basic observable is free of any theoretical bias, although we can
perform the calculation under the current theoretical framework. We
present numerical calculations at both HERMES kinematics for the
proton target and JLab kinematics for the neutron target. We find
that with the current theoretical understanding, Sivers effect plays
a key role in our analysis.
\end{abstract}

\begin{keyword}
left-right asymmetry, pion production, spin, semi-inclusive deep
inelastic scattering

 \PACS 13.60.Le, 13.85.Ni, 13.87.Fh, 13.88.+e
 \\ \center{To appear in PLB.}
\end{keyword}

\end{frontmatter}

\newpage

\section{Introduction}
Single spin asymmetries~(SSAs) on a transversely polarized target
provide us with rich information on the spin structure of the
nucleon, especially on the transverse spin. However, there has been
a prejudice that all transverse spin effect should be suppressed at
high energies in the past. It was not until in the 1990s, when the
E704 Collaboration reported their observation of a left-right
asymmetry in $p^\uparrow p\rightarrow\pi X$ process~\cite{E704},
that people began to show enthusiasms on transverse spin effects. In
order to account for the asymmetry, Sivers~\cite{Sivers} suggested a
possible mechanism, which is now called ``Sivers effect'',
originating from the asymmetry of the distribution function. But
this idea was criticized by Collins~\cite{Collins} on the ground of
violating the time reversal invariance of QCD. In
Ref.~\cite{Collins,Collins2}, another possible explanation, that
asymmetry arises from a fragmentation which is now known as
``Collins effect'',  was proposed. However, in Ref.~\cite{ssa1}, it
was argued that Sivers asymmetry might be allowed, and a good
description of E704 experiment was obtained by a parametrization. In
Ref.~\cite{ssa2}, another good description of E704 data was
obtained, but based on the Collins effect this time with a
surprising large contribution from unfavored fragmentation. Remember
that in Ref.~\cite{Ma}, the calculation is not so good to reproduce
the data based on Collins effect with the naive assumption of
favored fragmentation dominance. Later, the suppression of Collins
mechanism is also reproduced in Ref.~\cite{Ans22} by incorporating
the intrinsic partonic motion together with correct azimuthal
angular dependence. Now we have learnt~\cite{Alesio, Anselmino} that
there are three possible mechanisms contributing to the $p^\uparrow
p\rightarrow\pi X$ process: the Sivers effect, the Collins effect
and the Boer-Mulders effect~\cite{Boer-Mulders}. In
Ref.~\cite{Anselmino}, it was pointed that the Sivers effect is
important and other effects might be suppressed. We should also
aware that there is an alternative attempt to explain the left-right
asymmetry by the valence quark orbital angular moment
effect~\cite{Liang}, in distinct from the introduction of new
distribution and fragmentation functions.

Due to the complexity of the hadron-hadron process, we might as well
turn our point to a simpler process, the semi-inclusive deep
inelastic scattering~(SIDIS) process, which has attracted many
interests in recent years. Meanwhile, many progresses have been made
by experiments, e.g., non-vanishing SSAs have been observed by
HERMES~\cite{hermes} and COMPASS~\cite{compass} collaborations. On
the theoretical side, we have known that both Sivers and Collins
effects may contribute to the asymmetry. By multiplying different
weighting functions, the two effects can be separated, which is now
the conventional way of analyzing the data. Nevertheless, the
selection of the weighting functions strongly shows our bias on the
current theory. So in this paper, we will analyze the basic quantity
of left-right asymmetry in SIDIS process, following the analyzing
method by the E704 experiment, in which no weighting functions were
multiplied, to see whether a non-zero asymmetry can be obtained.
With the current theoretical knowledge, we find that the Sivers
effect plays a key role is our numerical calculation and indeed
produces a sizable left-right asymmetry in $\pi^{\pm}$ production
process. Therefore we suggest to measure the left-right asymmetries
in SIDIS process, for the purpose to provide a basic observable for
theoretical studies.

\section{Definition of the asymmetry}
In the E704 experiment~\cite{E704}, the left-right asymmetry is
defined as:
\begin{eqnarray}
A=-\frac{1}{P_B
\langle\cos\phi\rangle}\frac{N_\uparrow(\phi)-N_\downarrow(\phi)}{N_\uparrow(\phi)+N_\downarrow(\phi)}.
\end{eqnarray}
$P_B$ is the beam polarization and $\phi$ is the azimuthal angle
between the beam polarization direction and the normal to the
$\pi^\pm$ production plane. $N_{\uparrow(\downarrow)}$ is the number
of pions produced for beam spin tagged as positive (negative)
normalized to the beam flux.

Following the similar method, we define our asymmetry for the SIDIS
process as:
\begin{eqnarray}
A(\psi_s)=\frac{1}{S_T}\frac{N(\psi_s)-N(\psi_s+\pi)}
{N(\psi_s)+N(\psi_s+\pi)}=\frac{1}{S_T}\frac{d\sigma^\uparrow-d\sigma^\downarrow}
{d\sigma^\uparrow+d\sigma^\downarrow}.
\end{eqnarray}
$S_T$ is the transverse polarization of the target; $\psi_s$ is the
azimuthal angle between the transverse spin vector plane (defined by
spin vector and the incident beam) and a definite plane. The
definite plane can be chosen arbitrarily, e.g., we can choose the
horizontal plane in the laboratory frame for convenience. If
integrating the cross sections in the numerator and denominator
separately, we can investigate the asymmetry depending on various
kinematical variables.

Here we emphasize our difference with the conventional treatment.
When we perform the integration, no weighting functions are
multiplied, so we cannot integrate the azimuthal angles for the
produced hadrons from 0 to $2\pi$, which must lead to a vanishing
result. Instead, we will limit the azimuthal angles in a certain
range, e.g., $-\frac{\pi}{4}$ to $\frac{\pi}{4}$~(or
$\frac{3\pi}{4}$ to $\frac{5\pi}{4}$), i.e., only the hadrons
produced in a range to the left~(right) of the spin plane will be
selected, which is the way E704 experiment dealt with the data. This
detected region changes from left~(right) to right~(left) as the
target spin changes from up to down, thus a left-right asymmetry is
obtained. However, we have two choices to define the spin plane. In
E704 experiment, this plane was defined by the incident beam and the
spin vector, but in our paper, this plane is defined by the virtual
photon and the spin vector. We believe this is reasonable and
acceptable, for the DIS process can be considered as a virtual
Compton scattering process. So for the convenience of theoretical
description, the direction of the virtual photon is chosen as the
$z$-axis, which is denoted as the $\gamma^*p$ frame.
Correspondingly, $\ell p$ frame denotes the frame where the lepton
beam is defined as the $z$-axis. We can transform from one
coordinate system to another via a rotation by the angle $\theta$
between the exchanged photon and the incident beam. We
have~\cite{Diehl}:
\begin{eqnarray}
\sin\theta=\gamma\sqrt{\frac{1-y-\frac{1}{4}y^2\gamma^2}{1+\gamma^2}},~~~~~
\gamma=2xM_p/Q .
\end{eqnarray}
If $x$ is small, this angle is also small, which means that the
incident beam and the virtue photon almost lay in the same
direction. We make a rough estimation for HERMES
experiment~\cite{hermes}: $\langle x \rangle=0.09,~\langle y
\rangle=0.54,~\langle z \rangle=0.36,~\langle Q^2
\rangle=2.41\textmd{GeV}^2$, thus we have
$\langle\sin(\theta)\rangle\approx0.073$, which is indeed very
small. But we should be careful that as $x$ increases, this angle
might not be ignored.

\section{Expressions of the cross sections}
Due to the existence of the angle $\theta$, the component of a
vector can be different in different frames. For a transversely
polarized target, the polarization direction is perpendicular to the
incident beam, so the spin vector does not have the parallel
component in the $\ell p$ frame. But in the $\gamma^*p$ frame, a
parallel component of the spin vector is projected along the
$z$-axis, which means that we have longitudinal effect here although
the target is transversely polarized. By taking into account this
factor, the cross section up to leading twist is given as
follows~\cite{Diehl,Bacchetta}:
\begin{eqnarray}
\frac{d\sigma}{dx dy d\phi^\ell_S dz d\phi^\ell_h dP_{h\perp}^2}
&=&\frac{\alpha^2}{2sx(1-\epsilon)}\frac{\cos\theta}{1-\sin^2\theta\sin^2\phi_s^\ell}\times\bigg\{
\mathcal{F}[f_1D_1]\nonumber\\
&-&\frac{S_T\cos\theta}{\sqrt{1-\sin^2\theta\sin^2\phi_s^\ell}}
\sin(\phi_h^\ell-\phi_s^\ell)\mathcal{F}\bigg[\frac{\hat{\textit{\textbf{h}}}
\cdot\textit{\textbf{p}}_\perp}{M_p}f_{1T}^\perp D_1\bigg]\nonumber\\
&-&\frac{S_T\cos\theta}{\sqrt{1-\sin^2\theta\sin^2\phi_s^\ell}}
\sin(\phi_h^\ell+\phi_s^\ell)\mathcal{F}\bigg[\frac{\hat{\textit{\textbf{h}}}
\cdot\textit{\textbf{k}}_\perp}{M_h}h_1H_1^\perp\bigg]\bigg\}\nonumber\\
&\equiv&d\sigma_{UU}+d\sigma_{Siv}+d\sigma_{Col},
\end{eqnarray}
where
\begin{eqnarray}
\epsilon=\frac{1-y-\frac{1}{4}y^2\gamma^2}{1-y+\frac{1}{2}y^2+\frac{1}{4}y^2\gamma^2},~~~~~
\hat{\textit{\textbf{h}}}\equiv\textit{\textbf{P}}_{h\perp}/|\textit{\textbf{P}}_{h\perp}|.
\end{eqnarray}
The angles $\phi_h^\ell$ and $\phi_s^\ell$ are defined as:
$\phi_h^\ell=\phi_h-\phi^\ell,~~\phi_s^\ell=\phi_s-\phi^\ell$, where
$\phi^\ell$ denotes the orientation angle of the lepton plane.
Notice here that all the angles appearing in the cross section are
defined in the $\gamma^*p$ frame. In the above formula, we use a
compact notation:
\begin{eqnarray}
\mathcal{F}[\omega fD]=\sum_a e_a^2\int d^2\textit{\textbf{p}}_\perp
d^2\textit{\textbf{k}}_\perp
\delta^2(\textit{\textbf{p}}_\perp-\textit{\textbf{k}}_\perp-\textit{\textbf{P}}_{h\perp}/z)
\omega(\textit{\textbf{p}}_\perp,\textit{\textbf{k}}_\perp)
f^a(x,p^2_\perp)D^a(z,z^2k^2_\perp),~~~
\end{eqnarray}
where $\omega(\textit{\textbf{p}}_\perp,\textit{\textbf{k}}_\perp)$
is an arbitrary function. The factors depending on $\theta$ before
relevant terms are due to the transformation from $\gamma^*p$ to
$\ell p$ frames.

First, we may change the integration variables from $d\phi^\ell_S
d\phi^\ell_h$ to $d\phi^\ell d\phi_h$, and we can perform the
integration over $\phi^\ell$. We notice that
\begin{eqnarray}
&&\sin(\phi_h^\ell-\phi_s^\ell)=\sin(\phi_h-\phi_s),\nonumber\\
&&\hat{\textit{\textbf{h}}}\cdot\textit{\textbf{p}}_\perp=p_\perp\cos(\phi_h-\phi_{p_\perp}),\nonumber\\
&&\hat{\textit{\textbf{h}}}\cdot\textit{\textbf{k}}_\perp=k_\perp\cos(\phi_h-\phi_{k_\perp}),
\end{eqnarray}
all of which are independent of $\phi^\ell$, but
\begin{eqnarray}
&&\sin(\phi_h^\ell+\phi_s^\ell)=\sin(\phi_h+\phi_s-2\phi^\ell),\nonumber\\
&&\sin\phi^\ell_S=\sin(\phi_S-\phi^\ell),
\end{eqnarray}
both of which depend on $\phi^\ell$. If we ignore the difference
between the $\gamma^*p$ and $\ell p$ frame, we have
$\sin\theta=0,~\cos\theta=1$. After integration over $\phi^\ell$,
only the Sivers effect survives, and all the other terms including
the Collins term vanish. With a more strict management, we will not
ignore $\theta$, but expand the factors in $\sin^2\theta$, then we
have:
\begin{eqnarray}
&&\frac{1}{2\pi}\int^{2\pi}_0 d\phi^\ell
\frac{1}{1-\sin^2\theta\sin^2\phi^\ell_S} =1+\frac{1}{2}\sin^2\theta+o(\sin^4\theta),\nonumber\\
&&\frac{1}{2\pi}\int^{2\pi}_0 d\phi^\ell
\frac{\sin(\phi^\ell_h-\phi^\ell_S)}{(1-\sin^2\theta\sin^2\phi^\ell_S)^{3/2}}
=\sin(\phi_h-\phi_S)(1+\frac{3}{4}\sin^2\theta+o(\sin^4\theta)),\nonumber\\
&&\frac{1}{2\pi}\int^{2\pi}_0 d\phi^\ell
\frac{\sin(\phi^\ell_h+\phi^\ell_S)}{(1-\sin^2\theta\sin^2\phi^\ell_S)^{3/2}}
=-\sin(\phi_h-\phi_S)(\frac{3}{8}\sin^2\theta+o(\sin^4\theta)).
\end{eqnarray}
We find that the Sivers effect is $o(1)$, but the Collins effect is
$o(\sin^2\theta)$, which means that it is suppressed by $1/Q^2$.
Generally, only the terms independent of $\phi^\ell$ are $o(1)$, and
all the other effects are suppressed by $1/Q^2$, so Sivers effect is
dominant in our analysis, which is coincident with the analysis in
Ref.~\cite{Anselmino}.

In our calculation, we select the produced hadrons within the range
$\frac{3}{4}\pi\leqslant\phi_h\leqslant\frac{5}{4}\pi$, the right
side of the spin plane. Also we can choose the left side, and it is
clearly the same as we can see from the expression of the cross
section. Finally, we write the asymmetry for our numerical
calculation:
\begin{eqnarray}
A_{UT}(x,y,z)=\frac{\int d\phi^\ell_S dP^2_{h\perp} d\phi^\ell_h
~(d\sigma_{Siv}+d\sigma_{Col})}{\int d\phi^\ell_S dP^2_{h\perp}
d\phi^\ell_h ~d\sigma_{UU}}.
\end{eqnarray}

\section{Numerical calculations}

\begin{figure}
\center
\includegraphics[scale=0.7]{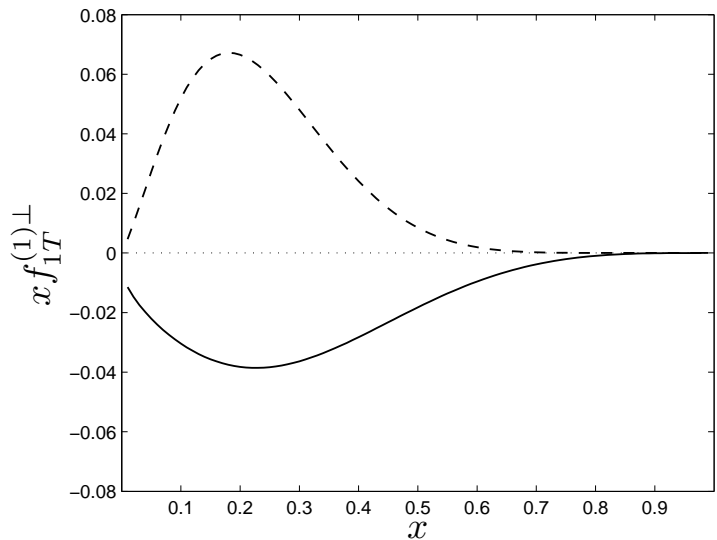}
\caption{$xf^{\perp(1) q}_{1T}(x)$ for $u$ and $d$ quarks in a
proton. The solid and dashed curves correspond to $u$ and $d$ quarks
respectively.} \label{siv1}
\end{figure}

To perform the calculation, we first need an input of Sivers
functions. However, there could be non-universality of transverse
momentum dependent distributions in different
processes~\cite{Bom08}, e.g., the Sivers asymmetry may enter in
hadron process with specific factors rather than simply a sign
change from SIDIS process. Therefore we should be cautious to apply
the parametrization extracted from one process to other kind of
processes~\cite{Bog99}. Fortunately, what we will calculate is for
the SIDIS process, and the parametrization of Sivers functions is
also from SIDIS data in Ref.~\cite{Anselmino_siv1, Anselmino_siv2},
in which the Sivers function is parameterized in the form:
\begin{eqnarray}
&&f^{\perp q}_{1T}(x,p^2_\perp)=-\frac{M_p}{p_\perp}\mathcal{N}_q(x)
f_q(x)g(p^2_\perp)h(p^2_\perp),\\
&&\mathcal{N}_q(x)=N_q x^a_q (1-x)^b_q
\frac{(a_q+b_q)^{(a_q+b_q)}}{a_q^{a_q} b_q^{b_q}},\\
&&g(p^2_\perp)=\frac{e^{-p^2_\perp/\langle p^2_\perp \rangle}}{\pi
\langle p^2_\perp
\rangle},~~~h(p^2_\perp)=\sqrt{2e}\frac{p_\perp}{M'}e^{-p^2_\perp/\langle
M'^2 \rangle}.
\end{eqnarray}
$f_1(x)$ is the unpolarized parton distribution functions, and we
adopt the CTEQ6L parametrization~\cite{CTEQ6} as an input. We plot
$f^{\perp(1) q}_{1T}(x)$, the one-moment of the Sivers function in
Fig.~\ref{siv1}.
This parametrization seems to indicate that $|f^{\perp(1)
d}_{1T}(x)| > |f^{\perp(1) u}_{1T}(x)|$, so we expect a larger
asymmetry in a neutron target than that in a proton target.

The fragmentation functions are~\cite{Kretzer}:
\begin{eqnarray}
D_{\mathrm{fav}}(z)&=&0.689z^{-1.039}(1-z)^{1.241},\nonumber\\
D_{\mathrm{unf}}(z)&=&0.217z^{-1.805}(1-z)^{2.037}. \label{frag}
\end{eqnarray}

In our calculation, we will consider the Collins effect, but as we
argued before that Collins effect is suppressed in our analysis, so
we will not care about the details on transversity and the Collins
functions, which are not known clearly yet. We will use the SU(6)
quark-diquark model~\cite{diquark} by including the Melosh-Wigner
rotation effect~\cite{ma} to describe transversity and adopt the
parametrization of Collins functions given by
Ref.~\cite{Anselmino_col}.

The kinematical cuts used in the calculation are shown in
Table~\ref{kin}.

\begin{table}
\begin{center}
\caption{kinematics} \label{kin}
\begin{tabular}{|c|c|}
\hline\hline
~~~~~~HERMES~~~~~~&~~~~~~JLab~~~~~~ \\
\hline\hline ~~~~~~$s=51.7$GeV$^2$~~~~~~&~~~~~~$s=23.4$GeV$^2$~~~~~~ \\
\hline ~~~~~~$Q^2>1$GeV$^2$~~~~~~&~~~~~$Q^2>1$GeV$^2$~~~~~~\\
\hline ~~~~~~$W^2>10$GeV$^2$~~~~~~&~~~~~$W^2>4$GeV$^2$~~~~~~\\
\hline ~~~~~$0.023<x<0.4$~~~~~~&~~~~~~$0.05<x<0.55$~~~~~\\
\hline ~~~~~$0.1<y<0.85$~~~~~~&~~~~~~$0.34<y<0.9$~~~~~\\
\hline ~~~~~$0.2<z<0.7$~~~~~~&~~~~~~~$0.3<z<0.7~~~~~~$ \\
\hline
\end{tabular}
\end{center}
\end{table}

For the HERMES experiment, a proton target is assumed, while for the
Jefferson Lab~(JLab) experiment, a neutron target is assumed. We
will investigate the $x$ and $z$ dependence\footnote{The E704
experiment only showed the dependence on $x_F$, i.e. approximate $z$
here.} of the asymmetries for both $\pi^+$, $\pi^-$ and $\pi^0$
productions.

\begin{figure}
\center
\includegraphics[scale=1]{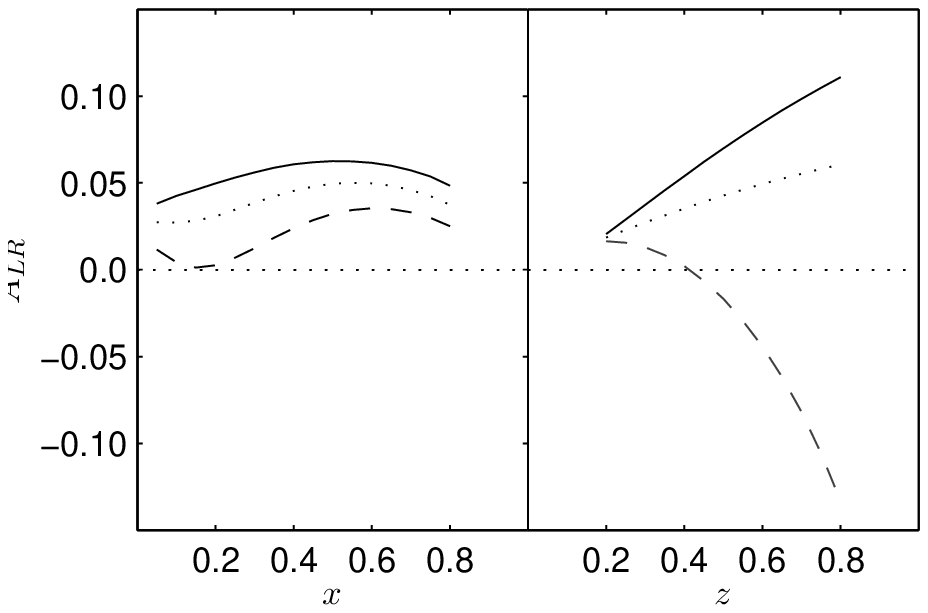}
\caption{Asymmetries for $\pi$ production at HERMES kinematics. The
solid, dashed and dotted curves correspond to the results for the
$\pi^+$, $\pi^-$ and $\pi^0$ production respectively. A proton
target is assumed here.} \label{hermes}
\includegraphics[scale=1]{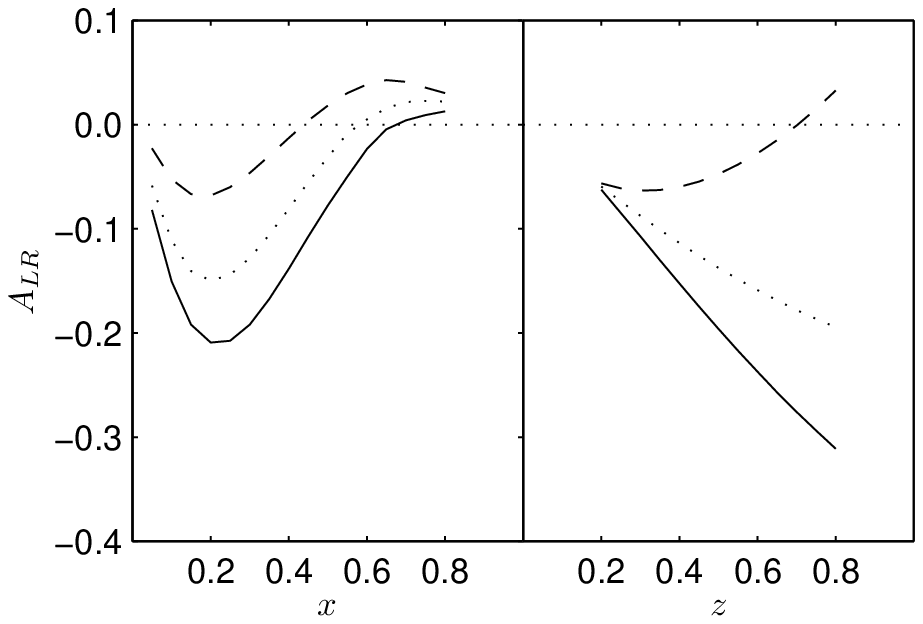}
\caption{The same as Fig.~\ref{hermes}, but a neutron target is
assumed here.} \label{jlab}
\end{figure}

Fig.~\ref{hermes} shows the results for $\pi$ production on a
transversely polarized proton target at HERMES kinematics, and
Fig.~\ref{jlab} shows the same results, but on a transversely
polarized neutron target at JLab kinematics. From these figures, we
clearly show non-vanishing asymmetries depending on $x$ and $z$.
Firstly, we notice that the asymmetries for $\pi^+$, $\pi^-$ and
$\pi^0$ productions are different, especially for the $z$-dependence
of the asymmetry, which is quite similar to that in the E704
experiment. This can be accounted for by different fragmentation
functions for different meson production. Secondly, the result for a
neutron target behaves completely different, almost opposite to that
for a proton target. If we notice that the Sivers functions for $u$
and $d$ quarks are of different signs, this can be deduced directly
from the isospin symmetry between the proton and the neutron. The
parametrization we used indicates that the Sivers distribution for
$d$ quarks is a little larger than that for $u$ quarks, thus a
larger asymmetry is obtained in a neutron target as the figures have
shown. However, we should be careful about it, and the correctness
of the parametrization needs a further check.

\section{Conclusion}
Single spin asymmetry~(SSA) is a powerful instrument to explore the
internal structure of the nucleon. A lot of theoretical works have
tried to obtain the asymmetries, and under the guidance, recent
experiments reported their discovery of the asymmetries. According
to the conventional treatment, various weighting functions should be
multiplied to project out the corresponding asymmetries. However,
the choice of a weighting function strongly shows a bias on a
certain theory, e.g., the current parton model based on operator
product expansion (OPE) and factorization. We do not consider it a
natural way dealing with the data, and it may not work if the theory
changes. In fact, there exist other theories such as the
recombination model~\cite{QRC1,QRC2} which can explain the spin
structure of the nucleon and the SSA phenomena. We expect a
``universal'' observable independent of any theory, and fortunately,
E704 experiment provided us an example.

In this paper, we analyzed the SIDIS process, following the method
by the E704 experiment. Our result clearly showed a left-right
asymmetry, with no weighting functions multiplied. Under the current
theoretical framework, we found that Sivers effect plays the key
role in our analysis, which might be helpful to understand the E704
experiment. We should emphasize that although our calculation
depends on the current theory, the basic observable of left-right
asymmetry is free of bias on any theories or models. We give the
predictions at both HERMES and JLab kinematics, and we suggest that
relevant experimental collaborations deal with their data in this
way to provide more information for theoretical studies.

\section*{Acknowledgement}
We acknowledge the valuable discussions with Erkang Cheng and Bo
Sun. This work is partially supported by National Natural Science
Foundation of China (Nos.~10721063, 10575003, 10528510), by the Key
Grant Project of Chinese Ministry of Education (No.~305001), by the
Research Fund for the Doctoral Program of Higher Education (China).

\end{document}